\documentclass[12pt]{iopart}
\usepackage{graphicx}
\usepackage{amssymb}
\usepackage{color}
\usepackage{amstext}
\usepackage{latexsym}
\usepackage[colorlinks,linkcolor=blue]{hyperref}

\def\la{\langle}
\def\ra{\rangle}
\def\om{\omega}

\def\vep{\varepsilon}

\newcommand{\beqa}{\begin{eqnarray}}
\newcommand{\eeqa}{\end{eqnarray}}

\begin{document}

\title{Causality and non-equilibrium second-order phase transitions in inhomogeneous systems}
\author{A. del Campo$^{1,2}$, T. W.  B. Kibble$^3$, W. H. Zurek$^1$} 
\address{$^1$Theoretical Division,  Los Alamos National Laboratory, Los Alamos, NM 87545, USA}
\address{$^2$Center for Nonlinear Studies,  Los Alamos National Laboratory, Los Alamos, NM 87545, USA}
\address{$^3$Blackett Laboratory, Imperial College, London SW7 2AZ, UK}

\begin{abstract}
When a second-order phase transition is crossed at fine rate, the evolution of the system stops being adiabatic as a result of the critical slowing down in the neighborhood of the critical point.
In systems with a topologically nontrivial vacuum manifold, disparate local choices of the ground state lead to the formation of topological defects. 
The 
universality class of the transition imprints a signature on the resulting density of topological defects: It obeys a power law in the quench rate,  with an exponent dictated by a combination of the critical exponents of the transition.
In inhomogeneous systems the situation is more complicated, as the spontaneous symmetry breaking competes with bias caused by the influence of the nearby regions that already chose the new vacuum.
As a result, the choice of the broken symmetry vacuum may be inherited from the neighboring regions that  have already entered the new phase. This competition between the inherited and spontaneous symmetry breaking enhances the role of causality, as the defect formation is restricted to a fraction of the system where the front velocity surpasses the relevant sound velocity and phase transition remains  effectively homogeneous.  
As a consequence, the overall number of topological defects can be substantially suppressed. When the fraction of the system is small, the resulting total number of defects is still given by a power law related to the universality class of the transition, but exhibits a more pronounced dependence on the quench rate. This enhanced dependence complicates the analysis but may also facilitate experimental test of defect formation theories.
\end{abstract}



\maketitle

\section{Introduction}

A phase transition is a transformation between different states of matter, and is typically induced by the quench of an external control parameter 
$\lambda$ through a critical point $\lambda_c$.
The Kibble-Zurek mechanism (KZM) is a theory designed to describe  the non-equilibrium dynamics in a scenario of spontaneous symmetry breaking and was originally developed for classical and continuous phase transitions \cite{Kibble76,Kibble80,Zurek85,Zurek93,KibblePT}.

The original question of what happens in rapid phase transitions arose in cosmology, where it was realized that, as a result of relativistic causality, essentially all field theories predict formation of topological defects, with dramatic consequences for the subsequent evolution of the Universe \cite{Kibble76,Kibble80}. It was later proposed that a similar mechanism of topological defect formation should occur in all phase transitions traversed at the finite rate (including condensed matter phase transitions), although (instead of relativistic causality) the resulting density of defects should be tied to the critical scalings \cite{Zurek85,Zurek93}: The critical slowing down means that the newly forming phase will be able to coordinate the choice of the symmetry breaking over domains of limited size.
We refer the reader to the earlier reviews \cite{Zurek96,Dziarmaga10,Polkovnikov11} of  the subject and recall the basics of the mechanism only briefly.
When a system is quenched through a critical point of a second-order phase transition both the equilibrium correlation length $\xi$
and relaxation time $\tau$ diverge according to
\beqa
\xi(t)  =  \frac{\xi_{0}}{|\varepsilon(t)|^{\nu}},
\qquad
\tau(t)  =  \frac{\tau_{0}}{|\varepsilon(t)|^{z\nu}},
\eeqa
in terms of the reduced parameter 
\beqa
\varepsilon=\frac{\lambda_c-\lambda}{\lambda_c},
\eeqa
 which varies from $\varepsilon<0$ ($\lambda>\lambda_c$) to $\varepsilon>0$ ($\lambda<\lambda_c$) as the transition is crossed from the high to the broken symmetry phase.
Above, $\nu$ and $z$ are the critical exponents that define the universality class of the transition.
The velocity with which the order-parameter phase information propagates (e.g.\ the speed of spin waves or second sound) is of the order of and no larger than the corresponding characteristic velocity
\beqa
s(t)=\frac{\xi(t)}{\tau(t)}=\frac{\xi_0}{\tau_0}|\vep|^{(z-1)\nu}.\label{s}
\eeqa

\begin{figure}
\begin{center}
\includegraphics[width=0.6\linewidth]{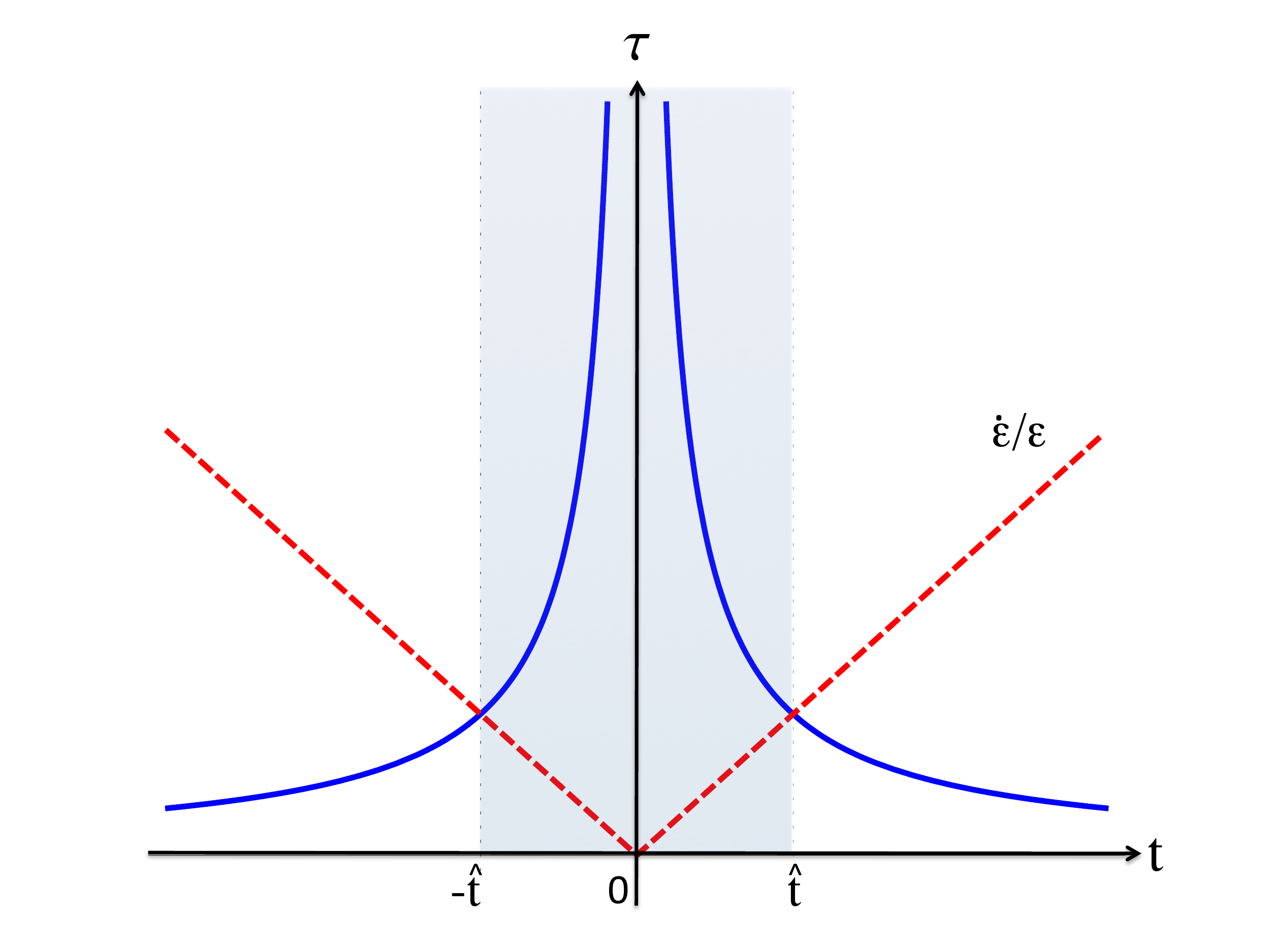}
\end{center}
\caption{\label{kzmplot} Impulse approximation for a second-order phase transition is central to the Kibble-Zurek mechanism. The equilibrium relaxation time $\tau$ diverges at the critical point, here reached at $t=0$ via a linear quench $\varepsilon=t/\tau_Q$. When the time remaining for the transition becomes comparable to the relaxation time, the system cannot follow the quench and the dynamics is no longer adiabatic -- it is no longer in equilibrium defined by the momentary value of $\varepsilon(t)$ (shaded area). The Kibble-Zurek mechanism estimates the typical size of the domains in the broken symmetry phase by the value of the equilibrium correlation length at the freezeout time $\hat{t}$.
}
\end{figure}

Assume a linear quench $\varepsilon(t)=t/\tau_Q$, which drives the system through the critical point at $t=0$.
 KZM simplifies the dynamics by distinguishing three stages: an adiabatic stage far away from the critical point, an impulse stage in the neighbourhood of the transition where the system is effectively frozen due to the divergence of the relaxation time, and a last stage governed by adiabatic dynamics again, sufficiently far away from the critical point and on the other (broken-symmetry) side of the transition. These three regimes are schematically represented in Fig.~\ref{kzmplot}. Within this picture, the freeze-out occurs at the instant $\hat{t}$ when the relaxation time of the system $\tau[\varepsilon(t)]$ 
equals to the time scale of the quench, $\varepsilon/\dot{\varepsilon}$. At that instant the dynamics approximately freezes (enters the impulse stage), and there is a breakdown of adiabaticity.  
Setting $\tau(t)\approx |\varepsilon/\dot{\varepsilon}|$, it follows \cite{Zurek85,Zurek93} that 
\beqa
\hat{t}\sim\left(\tau_{0}\tau_{Q}^{z\nu}\right)^{\frac{1}{1+z\nu}}.\label{eq:freeze_out_time}
\eeqa
The key insight of the KZM is that the average size $\hat{\xi}$ of the domains in the broken symmetry phase is set by the equilibrium correlation length evaluated at the freezeout time. This appears to hold even for transitions where only one side in Fig. \ref{kzmplot} 
is effectively present \cite{DMZ08,DTZ12}. 

A universal power-law scaling of the size $\hat{\xi}$ of domains that can coordinate spontaneous choice of broken symmetry on the quench rate emerges as a result, with the power given by a combination of the critical exponents,
\beqa
\hat{\xi}=\xi(\hat{t})=\xi_{0}\left(\frac{\tau_{Q}}{\tau_{0}}\right)^{\frac{\nu}{1+z\nu}}.
\label{KZMlength}
\eeqa
The symmetry in Fig. \ref{kzmplot} suggests that freezing occurs at $t\approx-\hat{t}$, and that any dynamics in the subsequent impulse stage,
 for $t\in[-\hat{t},\hat{t}]$, can be disregarded. However, numerical studies with asymmetric quenches around the critical point indicate that both sides may matter, and suggest that 
the domain formation may be decided at $\hat t$ after the transition, at $t\approx+\hat{t}$. \cite{DMZ08,DTZ12,AGR06,SZD12}. The density of excitations resulting from the quench scales as $d\sim\hat{\xi}^{-D}$ where $D$ is the dimensionality of the system. For topological defects this means a single ``piece'' of the defects per domain of size $\hat{\xi}$.

The prediction of defect density based on Eq.~(\ref{KZMlength}) suggests that the experimental study of non-adiabatic dynamics across a second-order phase transition may be used to probe the universality class of a system. In particular, one might be able to use it to determine experimentally the dynamical exponent $z$ which is often more difficult to measure than $\nu$. 

The adiabatic-impulse approximation can be invoked to describe the dynamics of a Landau-Zener  transition: KZM scaling yields surprisingly accurate predictions \cite{Damski05}, and can be related to avoided level crossing scenarios \cite{DZ06} (e.g., so that the asymmetry between the two sides of the transition can be also probed).
The KZM is also applicable to quantum phase transitions (QPTs) -- abrupt transitions between the nature of the ground state of quantum many-body systems that occur at zero temperature as a consequence of the gradual change of its Hamiltonian \cite{DSZB02, Polkovnikov05, ZDZ05, Dziarmaga05}. The control parameter $\lambda$ can be now related to e.g. an external magnetic field or to the strength of the interactions. The transition is predicted to leave the system with the density of excitations given by KZM scalings (see  reviews \cite{Dziarmaga10,Polkovnikov11}).
The relaxation time is then governed by the inverse of the energy gap between the ground state and the first excited state. This gap closes according to 
$\Delta\sim|\lambda-\lambda_c|^{z\nu}$, and  the original reasoning that yields $\hat t$ and $\hat{\xi}$ can be repeated. In addition to the resulting expected value of the defect density the amount of entanglement \cite{CDRZ07} as well as the effectiveness of the many-body system as a decohering environment \cite{DQZ11} can be also deduced using KZM paradigm in at least some cases.

\section{The inhomogeneous Kibble-Zurek mechanism}

The Inhomogeneous Kibble-Zurek mechanism (IKZM) is an extension of the paradigmatic KZM that allows one to describe the dynamics of a phase transitions in systems under an inhomogeneous quench $\lambda({\bf r},t)$, or 
with a spatially varying critical point $\lambda_c({\bf r})$, or a combination of both scenarios.

The earliest study of the effect of an inhomogeneous quench \cite{KV97} was primarily in the context of defect formation in $^3$He.  Experiments had been performed on phase ordering produced by thermal neutron injection into superfluid $^3$He-B \cite{Ruutu96, Bauerle96}.  When a neutron is absorbed via the exothermic reaction n + $^3_2$He $\to$ p + $^3_1$H + 0.76 MeV, a small volume is raised above the critical temperature and as it subsequently cools a network of vortex lines is formed.  Clearly the transition here is inhomogeneous; there is a thermal gradient and therefore the second-order transition occurs as a propagating front.  The key parameter is the velocity $v_F=(\tau_Q|\nabla\vep|)^{-1}$ of the front.   When it is large compared to the relevant sound velocity $\hat{s}$, the effect is very similar to the homogeneous case.  But when it is slow the phase of the order-parameter is constrained by that in the ordered state behind the front, so defect formation is restricted.  Because of critical slowing down, there is always a region near the front where $s < v_F$, where $s$ is given by Eq.~(\ref{s}), and therefore defects will form.  In the case of neutron-induced defect formation in $^3$He-B, it turns out that this region is of the same size as the bubble heated to above the critical point, so the homogeneous KZM estimate of defect density should still be reasonable, as indeed was confirmed by the experiments.  On the other hand, in $^3$He-A the transition is much slower and so fewer defects are formed.  Moreover, as we show below, the dependence of defect density on quench rate is much stronger in this case.
 
\begin{figure}
\begin{center}
\includegraphics[width=0.6\linewidth]{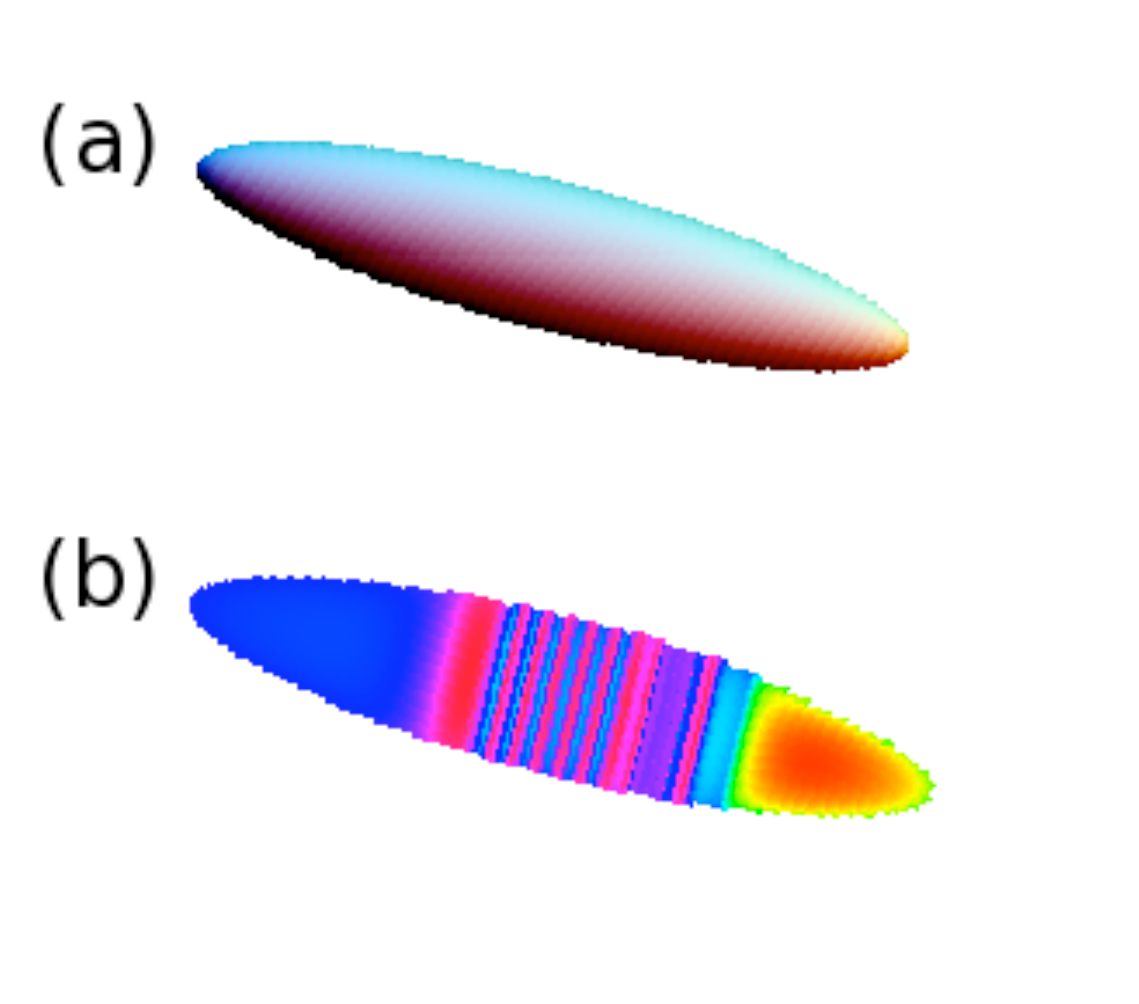}
\end{center}
\caption{\label{scheme}  Formation of solitons during the BEC induced by a thermal quench of a inhomogeneous atomic cloud: a) Isodensity contour of the gas in the harmonic trap. The critical temperatures for Bose-Einstein condensation depends on the density, and is highest in the centre of the trap. That is where the condensation will start. In sufficiently slow quench the phase selected by that central region would spread throughout the trap, resulting in a an equilibrium pool of BEC. b) However, for faster quenches, the choice of the ground state in the broken symmetry phase differs in different parts of the system and topological defects can form. Here, in the approximately one-dimensional ``cigar'' these are (semi-)topological {\it grey solitons} that correspond to sharp changes of the phase of the condensate. They will form only near the center of the trap, where the speed of the phase front exceeds the speed of the sound, and the transition is in effect homogeneous. As the phase front slows down near the ends of the ``cigar'', speed of sound may exceed the speed of the front, and the phase of the last region that selected the phase spontaneously will be then passed on (much like the preferred lattice orientation in the crystals grown by Czochralski process). Reprinted with permission from \cite{Zurek09}. Copyright 2009 American Physical Society.
}
\end{figure}

 The approach below was originally developed for the study of soliton formation in the course 
of Bose-Einstein condensation (BEC) \cite{Zurek09}. Gaseous BECs are confined in traps that result typically in a ``cigar'' shape (see Fig. \ref{scheme}) of the resulting condensate, which is approximately one-dimensional, and is densest in the center of the cigar (where BEC forms first). 

Let us consider the case when $\lambda_c=\lambda_c(x)$  (for simplicity we assume a one-dimensional system) and the external quench is homogeneous and linear in time, 
\beqa
\lambda(t)=\lambda_0\left(1-\frac{t}{\tau_Q}\right),
\eeqa 
for $t\in[-\tau_Q,\tau_Q]$. The choice $\lambda_0=\lambda_c(0)$ will be useful in the following (any other choice can be cast in this form by redefining $\tau_Q'=\tau_Q\lambda_c(0)/\lambda_0$).

The reduced control parameter becomes
\beqa
\label{rcp}
\varepsilon(x,t)=\frac{\lambda_c(x)-\lambda(t)}{\lambda_c(x)}.
\eeqa
This leads to an effective quench rate with a spatial dependence, 
\beqa
\tau_Q(x)=\left|\frac{\partial \varepsilon(x,t)}{\partial t}\right|^{-1}=\tau_Q\frac{\lambda_c(x)}{\lambda_c(0)}.
\eeqa
One can identify the location of the front $x_F$ crossing the transition at $t_F$ from the condition $\varepsilon(x_F,t_F)=0$. In particular, criticality is reached at point $x_F=x$ at a time,
\beqa
t_F(x)=\tau_Q\bigg[1-\frac{\lambda_c(x)}{\lambda_c(0)}\bigg].
\eeqa
This allows us to conveniently rewrite Eq.~(\ref{rcp})  as 
\beqa
\varepsilon(x,t)=\frac{t-t_F(x)}{\tau_Q(x)},
\eeqa
which resembles the expression in the homogeneous scenario.
We may pursue three alternative routes to identify the characteristic value of $\varepsilon(x,t)$ which determines the size of the domains in the broken symmetry phase.  
\begin{itemize}
\item
One way to go is to invoke the adiabatic-impulse-adiabatic approximation to determine the value $\hat{\varepsilon}(x,t)$ at which the system departs from the instantaneous equilibrium.
It suffices to match
\beqa
\tau[\varepsilon(x,t)]=\bigg|\frac{\varepsilon(x,t)}{\dot{\varepsilon}(x,t)}\bigg|=|\varepsilon(x,t)|\tau_Q(x).
\eeqa
It follows that 
\beqa
\hat{\varepsilon}(x)=|\varepsilon(x,\hat{t})|=\bigg[\frac{\tau_0}{\tau_Q(x)}\bigg]^{\frac{1}{1+\nu z}}.
\eeqa
\item
As an alternative, one can identify $\hat{\varepsilon}(x)$ by looking at the rate of change of the local equilibrium correlation length 
\beqa
s_{\xi}(\varepsilon)=\bigg|\frac{\partial \xi}{\partial \varepsilon}\frac{\partial \varepsilon}{\partial t}\bigg|=\frac{\nu\xi_0}{\tau_Q(x)|\varepsilon(x,t)|^{1+\nu}}.
\eeqa
required to match the velocity of perturbations, the 
relevant sound velocity  which can be upper bounded by the ratio of the local correlation length  and the relaxation time,
\beqa
s[\varepsilon(x)]=\frac{\xi_0}{\tau_0}|\varepsilon(x)|^{\nu(z-1)}.
\eeqa
Solving $s[\varepsilon(x,t)]=s_{\xi}[\varepsilon(x,t)]$ one finds $\hat{\varepsilon}(x)=\bigg[(\nu \frac{\tau_0}{\tau_Q(x)})\bigg]^{\frac{1}{1+\nu z}}$.  

\item A third alternative is based on the comparison of the local spatial extension of the sonic horizon $h(x)$ with the equilibrium correlation length.  Here 
\beqa
h[\varepsilon(x,t)]=\int _0^{T}s(T')dT'=\frac{s[\varepsilon(x,t)]\varepsilon(x,t)\tau_Q(x)}{\nu(z-1)+1}.
\eeqa
Solving $h[\varepsilon(x,t)]=\xi[\varepsilon(x,t)]$ we find $\hat{\varepsilon}(x)=\bigg[ [\nu(z-1)+1]\frac{\tau_0}{\tau_Q(x)}\bigg]^{\frac{1}{1+\nu z}}$. 
\end{itemize}

\begin{figure}
\begin{center}
\includegraphics[width=0.95\linewidth]{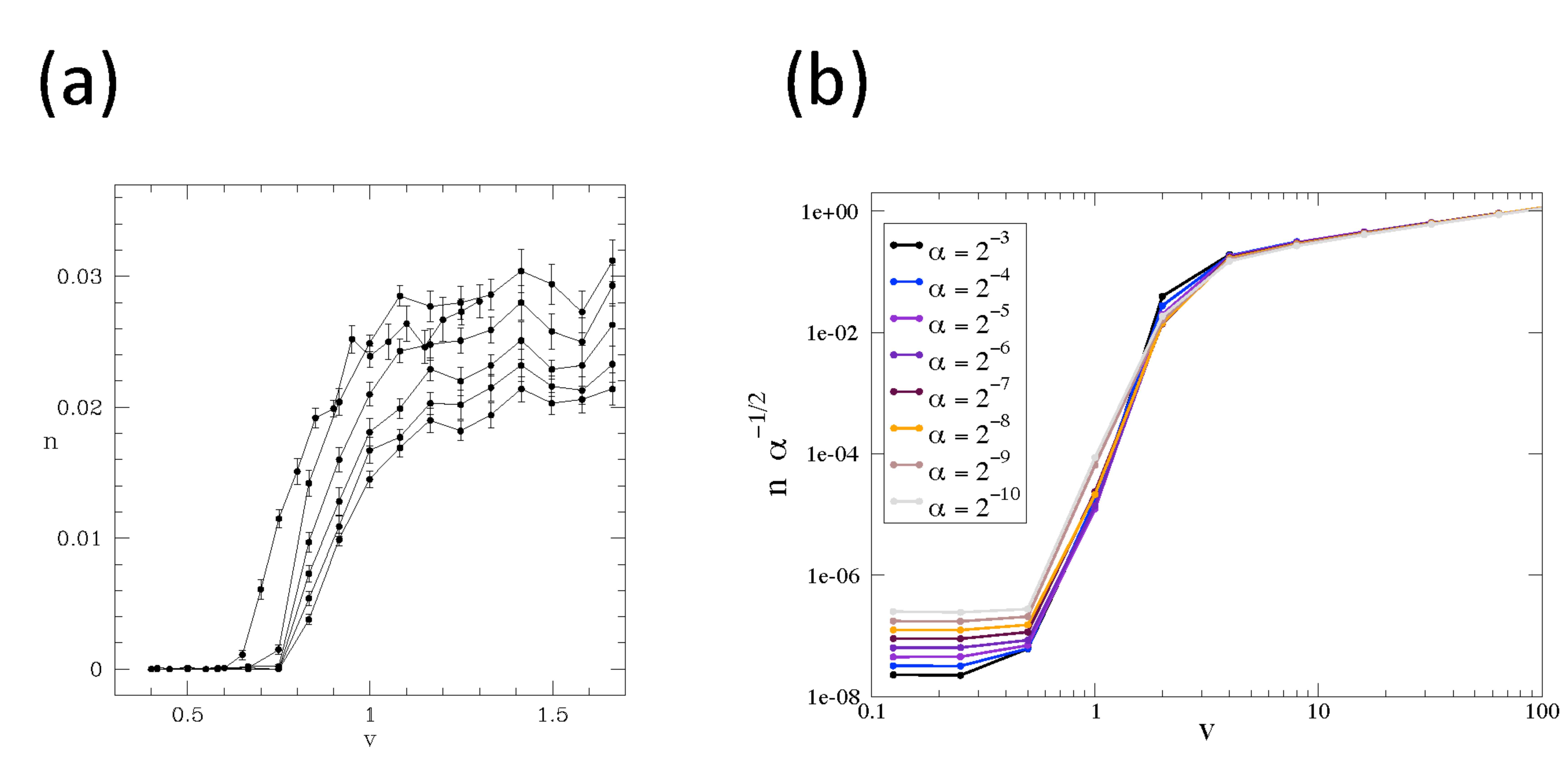}
\end{center}
\caption{\label{causality} Number of defects as a function of  the front velocity: a)  The simulations correspond to the dynamics of a one-dimensional real scalar field $\phi(x,t)$  governed by  a time-dependent Ginzburg-Landau equation of the form 
$\partial_t^2\phi+\eta\partial_t\phi-\partial_x^2\phi+[\phi^3-\epsilon(x,t)\phi]=\theta(x,t)$. Here, $ \theta(x,t)$ is a Gaussian white noise satisfying $\la\theta(x,t)\theta(x',t')\ra=2\eta\theta_0\delta(t-t')\delta(x-x')$ and the inhomogeneous quench  describes a sharp front $\epsilon(x,t)=\epsilon_0{\rm sgn}(t-x/v_F)$ propagating at velocity $v_F=v$.
The role of $\hat{s}$ is played by the characteristic sound velocity $s=[1+ 2\eta/3\sqrt{\epsilon_0}]^{-1/2}$, which takes the value $s=0.83$ for $\eta=\epsilon_0=1$. 
From top to bottom the curves correspond to  the temperature $\theta_0=10^{-1},10^{-2},10^{-4},10^{-6},10^{-8},10^{-10}$ and show that at low temperatures  the onset of excitations is well captured by the inequality $v_F>s$ \cite{DLZ99}. Copyright 1999 by the American Physical Society.
b) Formation of excitations in the inhomogeneous 1D quantum Ising Model described by the Hamiltonian  $H-=-\sum_{n=1}^Ng_n\sigma_n^x-\sum_{n=1}^{N-1}\sigma_n^z\sigma_{n+1}^z$. In the homogeneous case ($g_n=g$) 
in the thermodynamic limit, the system has critical points at $g=\pm1$ separating a paramagnetic phase ($|g|>1$) from a ferromagnetic phase ($|g|<1$).
The simulation corresponds to a chain of $N=400$ spins undergoing an inhomogeneous quench of the magnetic field of the form $g_n=1+\tanh[\alpha(n-vt)]$, 
driving locally the chain from the paramagnetic ($g=2$) to the ferromagnetic phase ($g=0$).  In this system $\hat{s}= 2$.
The formation of quasiparticle excitations is suppressed for $v<\hat{s}$ \cite{DM10}.
}
\end{figure}

Up to the prefactors given by the critical exponents, these alternative approaches agree on the prediction for $\hat{\varepsilon}(x)$,  associated with the local freezeout time interval
$[t_F(x)-[\tau_0 \tau_Q(x)^{\nu z}]^{\frac{1}{1+\nu z}},t_F(x)+[\tau_0 \tau_Q(x)^{\nu z}]^{\frac{1}{1+\nu z}}]$.
It follows that the typical size of the domains in the broken symmetry phase is given by
\beqa
\hat{\xi}(x)\simeq\xi[\hat{\varepsilon}(x)]=\xi_0\bigg[\frac{\tau_Q(x)}{\tau_0}\bigg]^{\frac{\nu}{1+\nu z}}.
\eeqa
This prediction is consistent with the use of  a local density approximation for $\tau_Q(x)$ in the KZM length scale predicted for homogeneous systems.
However, the role of causality is enhanced during the defect formation across a inhomogeneous phase transition. 
Defects are expected to be formed only where the velocity of  the front surpasses the characteristic velocity of perturbations $\hat{s}$ \cite{KV97,DLZ99}: in 
that case, the transition is effectively homogeneous, in that different fragments of the condensate cannot pass on their local choice of broken symmetry to their neighbours. Numerical and analytic studies in specific models \cite{SZD12,DLZ99,DM10,DM10b,Tanja12} have confirmed this expectation (see e.g. Fig. \ref{causality}).

The speed of propagation of this front can be estimated  to be
\beqa
v_F= \left|\frac {d x_F} {d t_F}\right| = \frac {\lambda_c(0)} {\tau_Q} \left| \frac {d \lambda_c(x)}{d x_F} \right|^{-1}.
 \label{eq:v_F}
\eeqa
As for the characteristic local velocity of the perturbations, it is given by the second-sound velocity  that can be upper bounded by the ratio of the local frozen out correlation length $\hat{\xi}(x)$
 and the relaxation time scale $\hat{\tau}(x)=\tau[\hat{\varepsilon}]=\hat{t}(x)$,
this is,  by 
\beqa 
\hat s =  \frac {\hat \xi} {\hat \tau} = \frac {\xi_0} {\tau_0} \bigg[\frac {\tau_0} {\tau_Q(x)} \bigg]^{\frac {\nu(z-1)} {1+\nu z}}.
\eeqa

The following cases can be distinguished. If $v_F<\hat s$ everywhere in the system, the resulting dynamics is adiabatic and defect formation is suppressed. As a result, IKZM may offer a way to an adiabatic dynamics across a second order phase transition.

The sharpness of this inequality has been investigated in a variety of models including the stochastic  Ginzburg-Landau  equation \cite{DLZ99}, as seen in Fig. \ref{causality}a.
It has also been studied in the 1D quantum Ising (see Fig. \ref{causality}b) and XY models \cite{DM10,DM10b}, the Gross-Pitaevskii equation for the miscible-immiscible phase transition of a two-component BEC \cite{SZD12}, and the Langevin dynamics of underdamped charged particles \cite{Tanja12}.
When $v_F>\hat s$ everywhere in the system, the dependence of the density of excitations on the quench rate is correctly described by the homogeneous KZM.
Clearly, this is always the case in homogeneous systems where  $v_F$ diverges given that $\lambda_c(x)=\lambda_c(0)$.

A novel scaling results when $v_F>\hat s$ holds only in  an effective fraction $\hat{X}$ of the system, 
depending explicitly on the quench rate.  
Depending on the topology of the system there might be several disconnected regions where this occurs \cite{DRP11}.
The number of excitations is then approximately given by
\beqa
d\simeq \frac{1}{L}\int_{\{x|v_F>\hat s\}} dx\frac{1}{\xi_0}\bigg[\frac{\tau_0}{\tau_Q(x)}\bigg]^{\frac{\nu }{1+\nu z}}.
\label{dikzm}
\eeqa
The condition $v_F>\hat s$ defines the effective size of the system where defect formation is allowed. Let us assume this is a simply connected region $[-\hat{x},\hat{x}]$.
Generally, the expression in Eq. (\ref{dikzm}) does not lead to a power-law  for $d$ in $\tau_Q$, and exhibits a more complex behavior dependence \cite{DRP11}.
 However, when the fraction $\hat{X}=\hat{x}/L\ll1$ (or more precisely, whenever $\lambda_c(x)\simeq\lambda_c(0)$ within $\hat{x}$) it is possible to derive analytical scaling power law. In several instances (including the case of Bose-Einstein condensation illustrated in Fig. \ref{scheme}) it has been found that \cite{Zurek09,ions1,ions2,DRP11}
\beqa
\label{ikzmscaling}
 d\simeq \frac{2\hat{x}}{\hat{\xi}(0)L}\propto \left(\frac{\tau_0}{\tau_Q}\right)^{\frac{1+2\nu}{1+\nu z}}.
\eeqa 
For mean-field critical exponents ($\nu=1/2$, $z=2$) this results in a power-law with $\frac{1+2\nu}{1+\nu z}=1$, four times larger than in the KZM where the exponent is $\frac{\nu}{1+z\nu}=1/4$. This pronounced  dependence on the quench rate is particularly amenable to tests in the laboratory.

 The IKZM applies as well to quantum phase transitions \cite{Dziarmaga10,Polkovnikov11,DM10,DM10b,ZD08}. The most important effect is similar to what we have seen in the classical phase transitions: Once the velocity of the front falls sufficiently below the local relevant sound speed, defect production is suppressed. Dziarmaga and Rams \cite{DM10,DM10b} have provided an especially convincing discussion of this phenomenon in the quantum Ising and XY models in one dimension.

\section{Laboratory systems}

\subsection{Trapped ions}

\begin{figure}
\begin{center}
\includegraphics[width=0.49\linewidth]{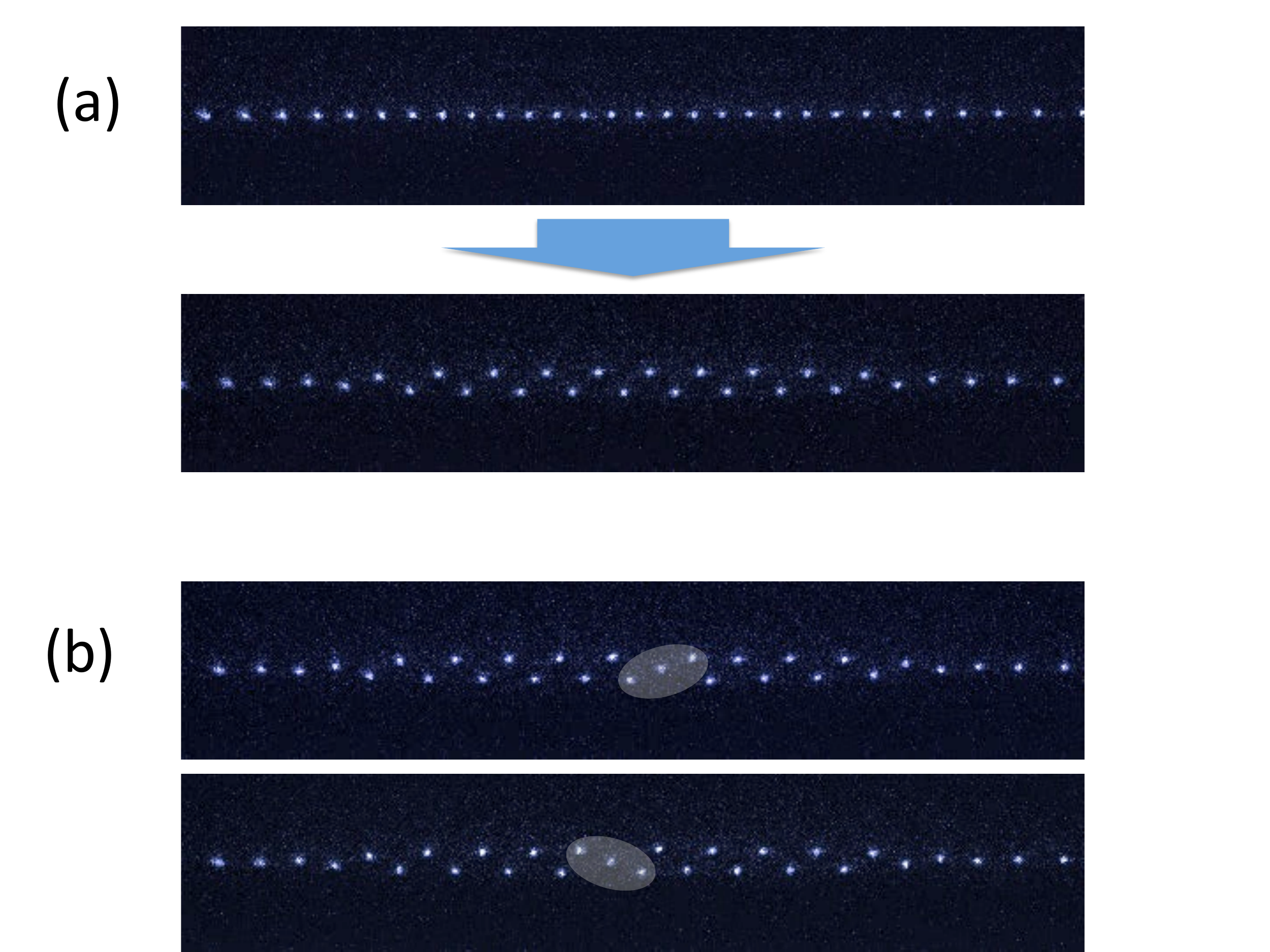}
\includegraphics[width=0.47\linewidth]{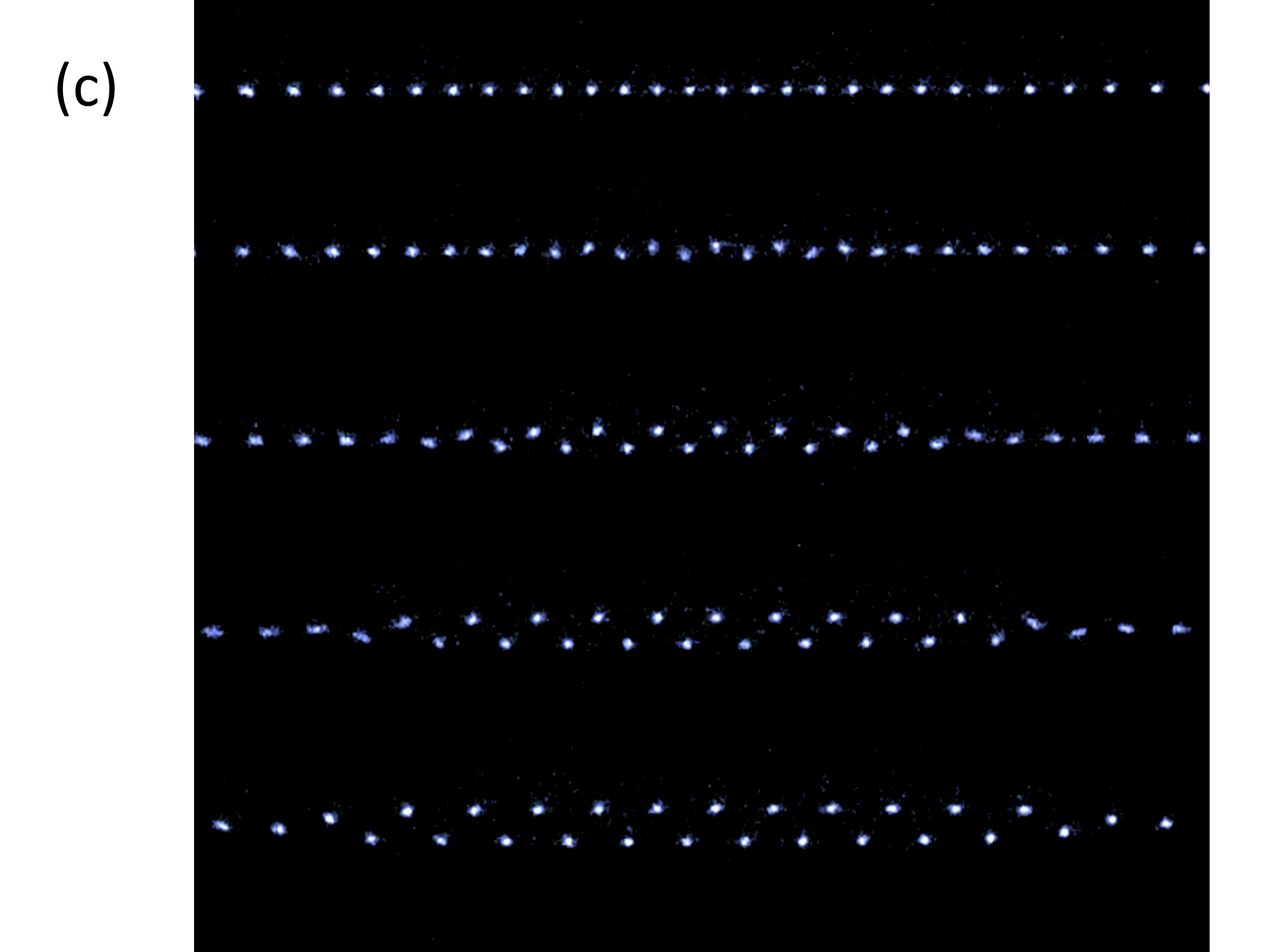}
\end{center}
\caption{\label{Tanja} (a) Structural phase transition of an ion Coulomb crystal in a harmonic trap. The chain consists of $30$ $^{172}{\rm Yb}^+$ ions in a linear Paul trap. As the transverse confinement is weakened there is a transition from the linear configuration to a doubly-degenerated zigzag chain.  (b) Snapshots of an ion chain supporting a kink and an anti-kink, identified in the shaded areas. (c) Sequence of snapshots showing the inhomogeneous nature of the transition. As a result of the longitudinal axial confinement the inter-ion spacing is inhomogeneous.  The transition is first crossed in the center of the chain, where the inter-ion Coulomb repulsion is higher. A front reaching the critical point spreads then sideways along the chain leading to the growth of the broken-symmetry zigzag phase. Courtesy of T. E. Mehlst\"aubler.
}
\end{figure}

 The quest for experimental evidence supporting the IKZM has recently
been pursued exploring the dynamics across a structural phase transition in ion
Coulomb crystals.
Trapped ions constitute a versatile platform for simulating many-body physics \cite{SPS12}. 
The formation of structural defects in ion crystals by quenching the external potential  was proposed in \cite{ions1,ions2}, and experimentally reported for the first time in \cite{NewRefSchaetz}.
An ion chain confined in a highly anisotropic Paul trap undergoes a continuous phase transition from a linear to a (doubly degenerate) zigzag structure when the transverse trap frequency is quenched across the critical value, see Fig.~\ref{Tanja}a. 
The interaction potential  is given by
\beqa
V=\frac{1}{2}m\sum_{j=1}^N \big[\nu_{||}^2x_j^2+\nu_{\perp}^2(y_j^2+z_j^2)]+\frac{1}{2}\frac{e^2}{4\pi\epsilon_0}\sum_j\sum_{j\neq k}\frac{1}{|{\bf r}_j-{\bf r}_k|},
\eeqa
for $N$ ions of mass $m$ and charge $e$.
The order parameter corresponds to the transverse size of the zigzag structure and evolves in accord with the Ginzburg-Landau potential \cite{Fishman08}.
Laser cooling of the ions provides 
the relaxation mechanism by which  vibrational excitations are damped. 
The possibility of testing the IKZM in this system arises from  the axial harmonic confinement
of frequency $\nu_{||}$, which induces a modulation of the linear ion density of the form  $n(x)=n(0)(1-x^2/L^2)$, where $L$ is half the length of the chain and $n(0)=\frac{3N}{4L}$. 
For homogeneous system, the critical value of the transverse confinement is given by $\nu_{\perp}\approx 2.051\om_0$ where the frequency characterizing the Coulomb coupling reads 
$\om_0=[\frac{e^2n(0)^3}{4\pi\epsilon_0m}]^{1/2}$.
Using a local density approximation, this  
critical frequency acquires a spatial modulation $\nu_{\perp}(x)=\nu_{\perp}[n(x)]$, making the phase transition  ultimately inhomogeneous, with the quadratic inhomogeneity analogous to the one illustrated in Fig.~\ref{scheme} (and, as in Fig.~\ref{scheme}, resulting from a harmonic trapping potential).
\begin{figure}
\begin{center}
\includegraphics[width=0.6\linewidth]{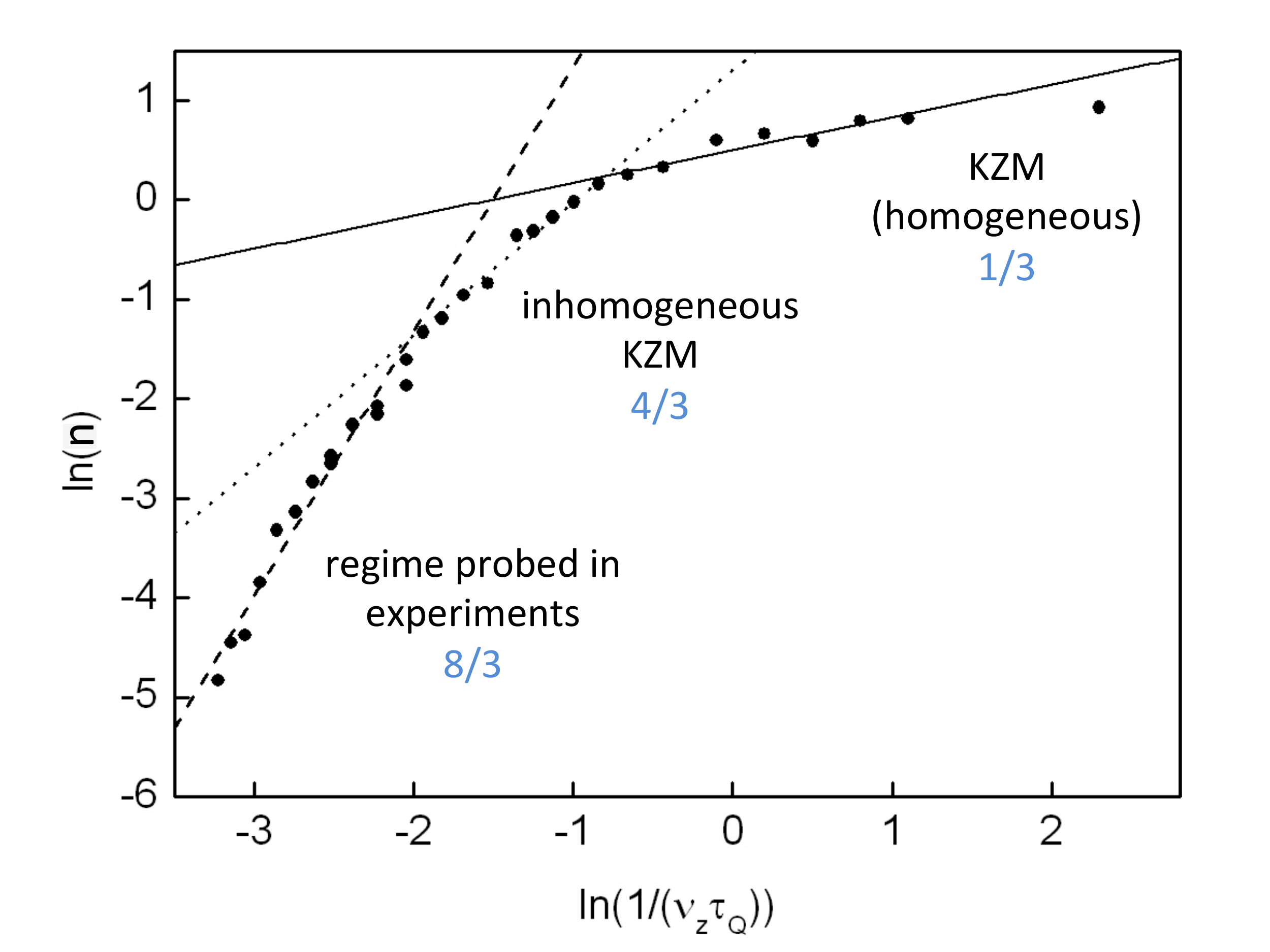}
\end{center}
\caption{\label{multiscaling} Scaling of the number of kinks in the quench rate for the linear to zigzag phase transition in an underdamped harmonically trapped ion chain. 
The different lines are guidelines to the eye illustrating the possible crossover among the different power laws anticipated. At fast quench rates (right part of the plot) the system behaves as affectively homogeneous and the standard KZM prediction applies. The power-law is described by the inverse of $\hat{\xi}$ in Eq. (\ref{KZMlength}) with an exponent $1/3$ for $\nu=1/2$, $z=1$. At intermediate quench rates causality determines the effective size of the system leading to a more pronounced dependence on the quench rate, as predicted in \cite{ions1,ions2}. In this regime the inhomogeneous Kibble-Zurek dictates a power-law exponent $4/3$. 
 At slower quench rates, the KZM correlation length $\hat{\xi}$
becomes comparable to the effectively homogeneous system size $\sim2\hat{x}$,
and typical realizations leads to one or no defects at all. The probability to find a single defect exhibits then a more pronounced dependence on the
quench rate, which can be approximately  fitted to a  power-law with exponent $8/3$ \cite{Tanja12}. This is the regime which would be probed experimentally in \cite{Tanja12,Ulm13,EH13}. 
 The mean kink number $n$ was obtained by averaging over 2000 realizations the number of kinks per cycle in a $N=30$ ion chain.
Adapted from  \cite{Tanja12}.}
\end{figure}
A non-adiabatic crossing results in the formation of structural defects, kinks and anti-kinks, as shown in Fig.~\ref{Tanja}b. Numerical simulations in \cite{ions1,ions2} show that this non-equilibrium dynamics is well described by the KZM. As the quench rate decreases, 
there is a crossover where among different 
power-laws, see Fig. \ref{multiscaling}.  For sufficiently fast quenches the whole ion chain is effectively homogeneous and the effect of the harmonic trap can be ignored. As a result the typical size of the zigzag domains scales in the quench rate in agreement with the KZM, i.e. Eq. (\ref{KZMlength}). As the quench rate decreases, it is possible to probe the inhomogeneous features of the system, and the scaling in Eq. (\ref{ikzmscaling}) governs the defect density.
The recent experiment at Physikalisch-Technische Bundesanstalt (PTB)  \cite{Tanja12} reported the scaling of defects as a function of the quench rate in an underdamped ion chain, which admits a description in terms of a Ginzburg-Landau theory with critical exponents $\nu=1/2$ and $z=1$.  
Typical realizations led to $\{0,1\}$ defects, as shown in Fig.~\ref{Tanja}. 
In that case of either $0$ or $1$ defects (of arbitrary topological charge) in the whole system, in a homogeneous phase transitions a doubling of the  KZM scaling was experimentally reported in \cite{Monaco06}. 
Subsequent theoretical works predicted a doubling of
the scaling in a variety of systems \cite{Saito07,DMZ08,Monaco09},
 although see \cite{Zureknew} for a more refined theory  of what doubles and what does not, which suggests a reexamination of the interpretation of the power laws observed in the experiments in tunnel Josephson junctions. Indeed, a  new
analysis of the experiment in \cite{Monaco06} has attributed the observed scaling to the onset of
an exponential suppression as opposed to a doubling of the KZM.
The PTB experiment suggests that a  doubling occurs as well in the IKZM. In particular, Fig. \ref{multiscaling} indicates that  the probability of forming a single defect $p_1$ scales as
\beqa
\label{dikzmscaling}
p_1 \simeq \bigg[\frac{2\hat{x}}{\hat{\xi}(0)}\bigg]^2\propto \left(\frac{\tau_0}{\tau_Q}\right)^{\frac{2(1+2\nu)}{1+\nu z}}=\left(\frac{\tau_0}{\tau_Q}\right)^{\frac{8}{3}},
\eeqa 
 where the last equality holds for $\nu=1/2$ and $z=1$.
Other experiments in ion chains of different sizes have reported  a similar scaling \cite{Ulm13,EH13}.
However, in this regime where $2\hat{x}\sim\hat{\xi}(0)$, signatures
of universality might be hidden by finite size-effects.
The limited range of available quench rates in these experiments does not
allow to conclusively elucidate this scenario.
Molecular dynamics simulations at lower quench rates show deviations
of the double power-law in Eq.~(\ref{dikzmscaling}).  
The current understanding of these experimental observations is
incomplete and the subject well deserves further studies.

\subsection{Bose-Einstein condensates}

Bose-Einstein condensation of trapped ultracold gases constitutes another relevant scenario where the inhomogeneous density of the cloud induced by the confinement  is expected to affect the non-adiabatic dynamics.
 Useful intuition about the transition can be gained by considering 
the Gross-Pitaevskii energy functional describing a trapped (scalar) Bose gas, which is given by
\beqa
\mathcal{E}^{\rm GP}[\Phi]=
\int \left(\frac{\hbar^2}{2m}|\nabla\Phi|^2+[V({\bf r})-\mu]|\Phi|^2+\frac{g}{2}|\Phi|^4\right) d^3{\bf r},
\eeqa
where $\mu$ is the chemical potential and $g$ in the interaction coupling.
For a homogeneous system the external potential can be set $V({\bf r})=0$, and Bose-Einstein condensation is expected whenever $\mu>0$. This transition corresponds to a $U(1)$ spontaneous symmetry breaking. In trapped systems, one can introduce the effective chemical potential $\mu({\bf r})=\mu-V({\bf r})$. 
Nonetheless, the measurements of the critical exponent $\nu$ indicate deviations from mean-field theory \cite{Esslinger07}.
\begin{figure}
\begin{center}
\includegraphics[width=0.6\linewidth]{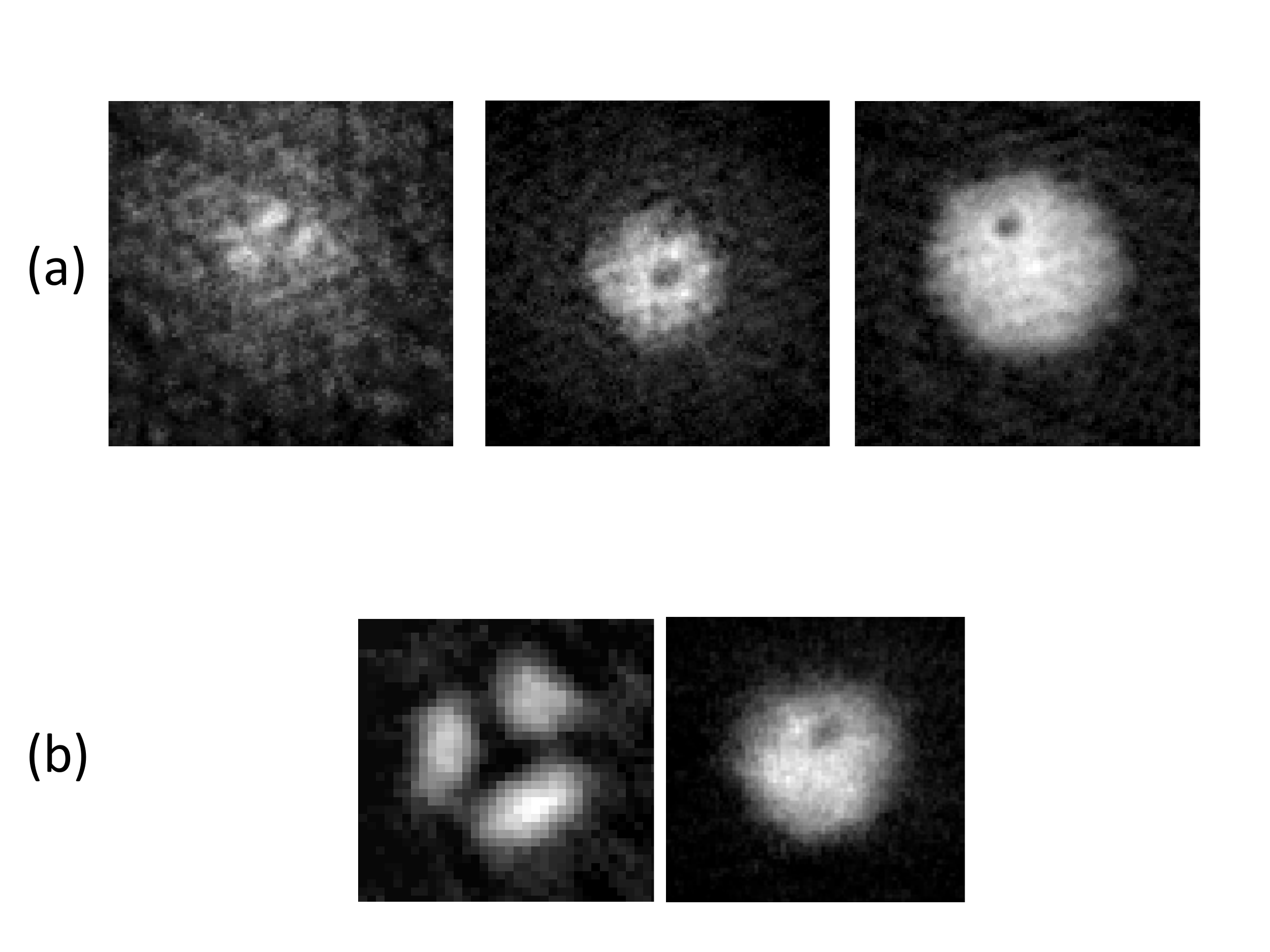}
\end{center}
\caption{\label{brianvortices} Spontaneous formation of vortices during the growth of a Bose-Einstein condensate. a) To illustrate the growth dynamics of a BEC in a 3D harmonic trap, snapshots of three different expanded BECs are obtained at different points in the BEC growth cycle.  Note that these images do not show the growth of a single BEC.  In left picture a few small regions of high density appear to stand out against a thermal-cloud background, although it is possible that these only appear as separated regions due to imaging artifacts.  The central picture shows a BEC in an early stage of growth, with a single clear vortex core and a few smaller dark regions that may represent additional vorticity.  In the snapshot at the right, a single vortex is clearly seen in a fully formed BEC. All images taken after expanding from the harmonic trap.
b) Merging of three independent Bose-Einstein condensates might lead to the spontaneous formation of a vortex, whenever the circulation of the phase around a given point adds up to a multiple of $2\pi$. Courtesy of B. P. Anderson.}
\end{figure}
At the University of Arizona at Tucson, 
experiments on forced evaporative cooling of a pancake-shaped thermal cloud  through Bose-Einstein condensation have demonstrated the spontaneous formation of vortices in harmonic and toroidal traps  \cite{Anderson08}. As the condensate starts to grow, the phase of the condensate wavefunction is chosen in patches of area $\sim\hat{\xi}(r)^2$ . As these patches merge with each other, there is a chance for the circulation of the phase around a given point to become a multiple of $2\pi$ and lead to the formation of a vortex.
This is illustrated in Fig. \ref{brianvortices}a where three different stages of the growth of a BEC are shown. Indeed,  the Tucson group 
 demostrated in a different experiment that the explicit merging of three independent BECs does lead to the formation of vortices \cite{Anderson07} (see Fig. \ref{brianvortices}b), with a  probability determined by the geodesic rule \cite{Andersonpra08}.
In \cite{Anderson08}, the limitation of the achievable quench rates and available statistics  prevented the test of the KZM scaling. It was later pointed out that the experimental setup  could explore both homogeneous and inhomogeneous scenarios of vortex formation \cite{DRP11}.

When the atomic cloud undergoing BEC is highly anisotropic, soliton creation follows from the non-adiabatic dynamics,  and the dependence of the number of spontaneously formed solitons on the quench rate observed in the homogeneous scenario \cite{DZ10} can be dramatically affected by the presence of the trapping potential \cite{Zurek09}. 
 Numerical simulations using the classical field method exhibit a power-law scaling \cite{WDGR11} in agreement with the IKZM scaling predicted in \cite{Zurek09}. 
Under similar conditions, soliton formation in a tight toroidal trap
has as well been studied in the context of spontaneous formation of
quantized circulation in a the newborn condensate  \cite{Das12}.
Tilting of the toroidal trap was shown to induce a suppression of the
number of solitons.
To assess the observability of IKZM during Bose-Einstein condensation, the above studies should be supplemented with a kinetic theory of evaporative cooling, relating the quench rate of the forced-evaporation to the effective cooling rate of the cloud. Preliminary results of an actual experiment \cite{newexp} yield scalings consistent with the theoretical predictions \cite{Zurek09}.

Inhomogeneities might be of relevance to the dynamics of other ultracold atom systems, as in the phase separation and pattern creation  in a binary Bose-Einstein condensate \cite{SZD11,SZD12},
the spontaneous spin vortex formation in a trapped ferromagnetic Bose-Einstein condensate \cite{SKU13},  or the quantum phase transition from  Mott insulator to superfluid phase, where a power-law scaling for the density of excitations has been experimentally demonstrated \cite{DeMarco11}.  
A universal scaling of the density of excitations governs as well the coherent quantum evolution of a many-particle system following a slow amplitude quench of a confining potential  \cite{CK10}.

In experiments with ultracold gases and trapped ions the spatial dependence of $\lambda_c(x)$ often arises as a result of the density modulation induced by an external trapping potential.
We close this section noticing that it is possible to introduce a trap-size scaling of the correlation length with the trap size 
\cite{CV09}.  This finite-size scaling modifies the divergence of the correlation length in the neighbourhood of the critical point. 
The discussion of the IKZM above is based on the power-law divergence of the equilibrium correlation length for a homogeneous system
and is therefore restricted to  quenches where the value of $\hat{\xi}$ is small compared to the trap size.

\section{Defect formation beyond scaling laws}

In this section we look for a more general expression for the formation of kinks, removing the restriction $\hat{X}\ll1$.  
We shall derive a general expression for the density of defects which cannot be cast as a power-law of the quench rate.
Nonetheless, we shall recover the results of the previous section for $\hat{X}\ll1$.

For the sake of illustration, we shall focus on systems where the spatial dependence of the critical control parameter is described by a Thomas-Fermi profile, i.e.
\beqa
\lambda_c(x)=\lambda_c(0)(1-X^2)^{\alpha},
\eeqa
where $\alpha\geq1$ and $X=x/L$. 
Particular cases correspond to the linear-to-zigzag transition in harmonically trapped ion chains \cite{ions1,ions2}, the mixing-demixing transitions \cite{SZD11,SZD12}, 
or the magnetic transition in spinor BEC \cite{SKU13}. A Gaussian spatial dependence of $\lambda_c(x)$ is of relevance to non-interacting ultracold gases and was discussed in \cite{Zurek09,DRP11}.
Consider a linear quench of the form 
$\lambda(t)=\lambda_c(0)(1-t/\tau_Q)$, for $t\in[-\tau_Q,\tau_Q]$.

 We note that  $\lambda_c(x)$ attains 
its maximum at $x=0$ so the critical point is first reached at the center. Symmetry breaking spreads sideways, 
and  the front crossing the transition reaches  point  $x$ at a time 
\beqa
t_F=\tau_Q\left(1-\frac{\lambda_c(x)}{\lambda_c(0)}\right).
\eeqa
Relative to this time, it is possible to rewrite
\beqa
\epsilon(x,t)=\frac{t-t_F}{\tau_Q(x)},
\eeqa
where the local quench rate is
\beqa
\tau_Q(x)=\tau_Q\frac{\lambda_c(x)}{\lambda_c(0)}=\tau_Q(1-X^2)^{\alpha}.
\eeqa
For the Thomas-Fermi profile the KZM bound to the velocity of perturbations reads
\beqa
\hat{s}(x)=\frac{\xi_0}{\tau_0}
\left(\frac{\tau_0}{\tau_Q}\right)^{\frac{\nu(z-1)}{1+\nu z}}
(1-X^2)^{-\frac{\alpha\nu(z-1)}{1+\nu z}},
\eeqa
while the front velocity is given by
\beqa
v_{\rm F}=\frac{L}{\tau_Q}\frac{1}{2\alpha|X|(1-X^2)^{\alpha-1}}.
\eeqa
Near the origin $X\approx 0$, $v_{\rm F}$ diverges and defect formation is expected.  Indeed, the inequality 
$v_{\rm F}(x_{\rm F}) >\hat s$ allows us to  find that defect formation is possible for all $x\in[-\hat{x},\hat{x}]$, where
\beqa
\label{hatxapprox}
\hat{x}=\frac{L^2}{2\alpha\xi_0}\left(\frac{\tau_0}{\tau_Q}\right)^{\frac{1+\nu}{1+\nu z}}+\mathcal{O}(\hat{X}^3).
\eeqa

The number of defects is given by
\beqa
d&\simeq &\frac{1}{L}\int_{-\hat{x}}^{\hat{x}} dx \frac{1}{\xi_0}\left(\frac{\tau_0}{\tau_Q(x)}\right)^{\frac{\nu}{1+\nu z}}\nonumber\\
&=&
\frac{2\hat{x}}{\hat{\xi}(0)L}\,\,{_2F_1}\left[\frac{\alpha\nu }{1+\nu z},\frac{1}{2},\frac{3}{2};\left(\frac{\hat{x}}{L}\right)^2\right],
\eeqa
where $_2F_1(a,b,c;z)$ is the hypergeometric function, provided that $\hat{x}\leq L$, as is the case. 
As an upshot, whenever the spatial dependence of the effective quench rate $\tau_Q(x)$ becomes important over the domain where defect formation is allowed by causality, 
the resulting number of kinks is not described by a power-law scaling any more.
Slow quenches and tight traps provide actually the ideal scenario to observe IKZM scaling, see \cite{DRP11}, end of Sec.~3 (assuming no decay mechanisms).
For $\hat{x}\ll L$, 
\beqa
\label{dexpansion}
d &\simeq& \frac{2\hat{x}}{\hat{\xi}(0)L}
\bigg[1-\frac{\alpha\nu}{3(1+\nu z)}\frac{\hat{x}^2}{L^2}+\mathcal{O}(\hat{X}^4)\bigg],
\\
&=&  \frac{1}{\alpha \xi_0}\left(\frac{\tau_0}{\tau_Q}\right)^{\frac{1+2\nu}{1+\nu z}}+\mathcal{O}(\hat{X}^3),
\eeqa
where we recognize the leading term 
as the estimate for the density of defects in the previous subsection.
Whenever $\hat{x}$ can be approximated by Eq.~(\ref{hatxapprox}), this term leads to a power-law scaling as a function of the quench rate. 
One can identify two sources of deviations from this scaling.
First, the exact solution of $v_{\rm F}(x_{\rm F}) >\hat s$  for $\hat{x}$, which typically leads to a dependence on $\tau_Q$ which is not  a power-law. As mentioned above,  it can be the case that  $v_{\rm F}(x_{\rm F}) >\hat s$ holds in spatially separated regions of the system leading to a more complicated dependence of $\hat{X}$ on $\tau_Q$ \cite{DRP11}.
Second, the local dependence of the quench rate $\tau_Q(x)$, which is responsible for the higher leading terms in the expansion in brackets, Eq.~(\ref{dexpansion}). 

So far, we have restricted the presentation to one-dimensional systems. However, the extension to higher dimensions is straightforward.
For the sake of illustration, consider a two dimensional systems with radial symmetry, area $A=\pi L^2$, and critical control parameter $\lambda_c(r)=\lambda_c(0)(1-R^2)^{\alpha}$ where $R=r/L$. The density of defects of size $\xi(t)^2$ is given by
\beqa
d&\simeq &\frac{1}{\pi L^2}\int_{0}^{\hat{r}} \frac{2\pi r dr}{\xi_0^2}\left(\frac{\tau_0}{\tau_Q(r)}\right)^{\frac{2\nu}{1+\nu z}}\nonumber\\
&=&\frac{1}{\xi_0^2}\left(\frac{\tau_0}{\tau_Q}\right)^{\frac{2\nu}{1+\nu z}}\frac{1-(1-\hat{R}^2)^{\beta}}{\beta}\nonumber\\
&=& d_{\rm KZM}f(\hat{R}).
\eeqa
Here, $\beta=1-\frac{2\nu\alpha}{1+\nu z}$ and we have assumed that the causality condition restricts the formation of defects to $r\in[0,\hat{r})$ and defined $\hat{R}=\hat{r}/L$.
This expression is particularly interesting since the term $d_{\rm KZM}=\frac{1}{\xi_0^2}\left(\frac{\tau_0}{\tau_Q}\right)^{\frac{2\nu}{1+\nu z}}$ can be recognized as the KZM estimate for the density of defects in a homogeneous system.
The role of the inhomogeneity is encoded in the additional term $f(\hat{R})$, which can be expanded whenever $\hat{R}\ll1$ to $f(\hat{R})=\hat{R}^2+\mathcal{O}(\hat{R}^4)$. Noting that
$\hat{r}$ can be approximated by the form of $\hat{x}$ in Eq. (\ref{hatxapprox}), we derive the power-law scaling
\beqa
d&=&\frac{L}{2\alpha \xi_0^3}\left(\frac{\tau_0}{\tau_Q}\right)^{\frac{2(1+2\nu)}{1+\nu z}}+\mathcal{O}(\hat{R}^4),
\eeqa
For example, this dependence could be experimentally probed  by studying vortex formation during  Bose-Einstein condensation  induced by a slow thermal quench in a tight trap \cite{DRP11}.

We note that in these higher dimensional systems there exist the possibility for the critical control parameter to be inhomogeneous only along a given set of degrees of freedom. This results in power-law scalings with exponents intermediate between those of the homogeneous and inhomogeneous scenarios. An example of relevance to the experiment  \cite{Anderson08}
is the spontaneous vortex formation during the growth of a Bose-Einstein condensation in toroidal trap \cite{DRP11}.

\section{Discussion: Inhomogeneous driving as a tool to suppress excitations}

Suppression of excitations across a phase transition is of interest in a wide variety of applications, including the preparation of quantum phases in the ground state in a quantum simulator \cite{QS82,CZ12}, and  implementing protocols for adiabatic quantum computation \cite{AQC}. Different approaches have been put forward to approach or mimic adiabaticity. 
Finite size-effects pave the way to an adiabatic dynamics, whenever the  $\hat{\xi}$ is larger than the system size $L$.
An adiabatic dynamics can be implemented exploiting  the energy gap that arises from the finite size of the system \cite{Murg04}.
An alternative approach is to consider a finite-rate non-linear quench  of the form $\lambda(t)=\lambda_0|t/\tau_Q|^r$ and optimize the power-law exponent $r$ to 
minimize the density of excitations following the quench through the critical point \cite{BP08,SSM08}. 
The optimal power turns out to be given by the logarithm of the quench time multiplied
by universal critical exponents characterizing the phase transition \cite{BP08}. However, an accurate knowledge of the critical point and excellent control of $\lambda(t)$  are essential in this approach. Otherwise, the quench might be linearized in the proximity of the critical point. 
The KZM scaling can be modified as well by coupling a quantum critical system to an environment. The resulting nonequilibrium dynamics exhibits corrections to the KZM scaling and paves the way to relaxation of excitations \cite{Patane08}.
In addition, the topology of the system can also influence production of defects. This fact was illustrated in the Creutz model, 
which supports topological edge states \cite{Bermudez09}. It was shown that 
the dynamics following a quench of an edge states dramatically increases the exponent governing the power-law scaling and diminishes the density of excitations.  This observation was extended in a subsequent study of the adiabatic dynamics of Majorana fermions across a quantum phase transition. It was shown that the KZM describes the formation of bulk defects but it is not valid for
for edge Majorana fermions.  The root of these non-universal deviations was pinned down in the localization dynamics of edge states \cite{Bermudez10}. 
Further advances in tailoring excitations have been achieved using optimal quantum control (OQC) strategies. The combination of time-dependent Density Matrix Renormalization Group (t-DMRG) with the Chopped RAndom Basis (CRAB) optimization algorithm allows to tackle a great variety of non-integrable quantum many-body one-dimensional systems \cite{OQC,OQC2}. Remarkably, OQC allows one to optimize adiabatic evolution with neither knowledge of the instantaneous eigenstates nor the resulting gap of an arbitrary time-dependent system \cite{OQC3}. Of course, the applicability of OQC in many-body systems is not restricted QPT, and can be applied to design quenches without crossing a critical point  \cite{RC11}.
Yet another approach, consists in using counterdiabatic interactions which allow to drive the dynamics through the instantaneous eigenstates of the many-body system \cite{DRZ12}.

The IKZM can be thought of  as a complementary tool to these various 
approaches since it helps to reduce the number of excitations generated during the crossing of a phase transition \cite{KV97}. Its relevance to many-body state preparation and adiabatic quantum computation was pointed out in \cite{DM10,DM10b}, see as well \cite{ZD08}. 
This line of thinking has recently been applied to the adiabatic growth of  an antiferromagnetic phase in a Mott insulator made of two atomic species \cite{GE12}.

\section{Conclusion}

The divergence of the relaxation time in the neighbourhood of the critical point induces a non-adiabatic dynamics in which causally disconnected regions perform independent choices of the broken-symmetry vacua. As result, topological defects are formed, their density scaling with the quench rate as described by the Kibble-Zurek mechanism.
In inhomogeneous systems, the local choice of the ground state in the broken symmetry phase can be communicated to neighbouring regions.
Criticality is reached locally as the phase transition front spreads gradually across the system. When its velocity is lower than the speed of the relevant sound the formation of defects is suppressed:  Otherwise causality determines the effective size of the system.
This competition between local, spontaneous choice of broken symmetry and the bias of the neighbouring regions that had already made their selection leads to a more pronounced dependence of the defect density on the quench rate than in the homogeneous scenario. 
The latter is only recovered in inhomogeneous systems for fast quench rates.
Regimes where the dependence is captured by a simple power-law have been identified, as well as a variety of crossovers among different power-laws.
This universal nonequilibrium dynamics  is of relevance to a wide variety of laboratory systems including trapped ions and ultracold gases.

\ack

It is a pleasure to thank Tanja E. Mehlstaubler for facilitating the snapshots of ion chains used in Fig. \ref{Tanja},  Brian P. Anderson for the pictures of growing Bose-Einstein condensates supporting vortices reported in Fig. \ref{brianvortices}, and Marek M. Rams for the numerical data in Fig. \ref{causality}b. We further thank Bogdan Damski and Marek M. Rams for useful suggestions,
and Emilie Passemar for help in editing the figures.
This research is supported by the U.S Department of Energy through the LANL/LDRD Program and a  LANL J. Robert Oppenheimer fellowship (AdC).

\section*{References}


\begin{thebibliography}{10}





\bibitem{Kibble76}  Kibble  T W B 1976 {\it J. Phys. A: Math. Gen.}  {\bf 9} 1387
\bibitem{Kibble80} Kibble  T W B  1980  
{\it Phys. Rep.} {\bf 67} 183

\bibitem{Zurek85}  Zurek W H 1985
{\it Nature (London)} {\bf 317} 505
\bibitem{Zurek93} Zurek W H 1993 {\it Acta Phys. Pol. B} {\bf 24} 1301

\bibitem{KibblePT}  Kibble  T W B 2007 {\it Physcis Today}  {\bf 60} 47

\bibitem{Zurek96} Zurek W H 1996 
{\it Phys. Rep.}  {\bf 276} 177


\bibitem{Dziarmaga10} Dziarmaga J 2010 
{\it Adv. Phys.} {\bf 59} 1063

\bibitem{Polkovnikov11} 
Polkovnikov A, Sengupta K, Silva A and Vengalattore M  2011 
{\it Rev. Mod. Phys.}  {\bf 83} 863

\bibitem{DMZ08} Dziarmaga J, Meisner J and Zurek W H 2008
{\it Phys. Rev. Lett.} {\bf 101} 115701

\bibitem{DTZ12}  Dziarmaga J, Tylutki M and Zurek W H 2012
{\it Phys. Rev.} B {\bf 86} 144521 

\bibitem{AGR06} Antunes N D,  Gandra P and Rivers R J 2006 
{\t Phys. Rev.} D {\bf 73} 125003 
 
\bibitem{SZD12} Sabbatini J,  Zurek W H, and Davis M J 2012 
 {\it New J. Phys.} {\bf 14} 095030

\bibitem{Damski05} Damski B 2005
{\it Phys. Rev. Lett.} {\bf 95} 035701 

\bibitem{DZ06} Damski B and Zurek W H  2006 
{\it Phys. Rev. A} {\bf 73} 063405

\bibitem{DSZB02} Dziarmaga J,  Smerzi A, Zurek W H and Bishop A R  2002 
{\it Phys. Rev. Lett.}  {\bf 88} 167001

\bibitem{Polkovnikov05} Polkovnikov A 2005 
{\it Phys. Rev. } B {\bf 72} 161201(R)

\bibitem{ZDZ05} Zurek W H, Dorner U and Zoller P 2005
{\it Phys. Rev. Lett.} {\bf  95} 105701

\bibitem{Dziarmaga05} Dziarmaga J 2005  
{\it Phys. Rev. Lett.} {\bf  95} 245701

\bibitem{CDRZ07}  Cincio L, Dziarmaga J, Rams M M and Zurek W H 2007 
{\it  Phys. Rev.} A {\bf 75} 052321

\bibitem{DQZ11}   Damski B, Quan H T and  Zurek W H 2011
{\it Phys. Rev.} A {\bf 83} 062104




\bibitem{KV97}  Kibble T W B and  Volovik G E 1997 {\it JETP Lett.} {\bf 65}  102

\bibitem{Ruutu96} Ruutu V M H,  Eltsov V B, Gill A J, Kibble T W B, Krusius M, Makhlin Yu G, Pla\c ais B, Volovik G E, Xu Wen 1996
{\it Nature} {\bf 382} 334

\bibitem{Bauerle96} B\"auerle C, Bunkov Yu M, Fisher S N, Godfrin H, Pickett G R 1996
{\it Nature} {\bf 382} 332

\bibitem{Zurek09} Zurek W H 2009 
 {\it Phys. Rev. Lett.} {\bf 102} 105702 

\bibitem{DLZ99} Dziarmaga J,  Laguna P, Zurek W  H 1999
 {\it Phys. Rev. Lett.} {\bf 82} 4749 

\bibitem{DM10} Dziarmaga J and  Rams M M 2010
{\it New J. Phys.}  {\bf 12} 055007 

\bibitem{DM10b}  Dziarmaga J and Rams M M 2010 
{\it New J. Phys.}  {\bf 12} 0103002 
 


\bibitem{Tanja12}  Pyka K, Keller J,  Partner H L,  Nigmatullin R,   Burgermeister  T,  Meier D M,  Kuhlmann K,  Retzker A,  
 Plenio M B,   Zurek W H, del Campo A and   Mehlst\"aubler T E {\textcolor{red} 2012  {\it Nat. Commun.} {\bf  4} 2291 }


\bibitem{DRP11} del Campo A, Retzker A and Plenio M B 2011 
{\it New J. Phys. } {\bf 13} 083022 


\bibitem{ions1} del Campo, A, De Chiara G, Morigi G, Plenio M B and Retzker A 2010
{\it Phys. Rev. Lett.}  {\bf 105} 075701

\bibitem{ions2} De Chiara G,  del Campo A, Morigi G, Plenio M  B and Retzker A 2010
 {\it  New J. Phys.}  {\bf  12} 115003 

\bibitem{NewRefSchaetz} Mielenz M {\it et al}  2013 {\it Phys. Rev. Lett.} {\bf  110} 133004

 \bibitem{ZD08}  Zurek W H and Dorner U  2008
{\it Phil. Trans. R. Soc. } A {\bf 366} 2953


\bibitem{SPS12} Schneider C, Porras D and Schaetz T 2012
{\it Rep. Prog. Phys.} {\bf  75} 024401


\bibitem{Fishman08} 
Fishman S, De Chiara G, Calarco T and Morigi G 2008 {\it Phys. Rev.} B {\bf 77} 064111


\bibitem{Monaco06} Monaco R, Mygind J, Aaroe M, Rivers R J and Koshelets V P 2006
 {\it Phys. Rev. Lett.} {\bf 96} 180604

\bibitem{Saito07} Saito H, Kawaguchi Y and Ueda M 2007 
{\it Phys. Rev.} A {\bf 76} 043613


\bibitem{Monaco09} Monaco R, 
Mygind J, 
Rivers R J and
Koshelets V P 
2009 
{\t Phys. Rev.} B {\bf 80} 180501(R).


\bibitem{Zureknew} Zurek W H 2013, arXiv:1305.4695


\bibitem{Ulm13}  Ulm S,  Ro\ss nagel J,  Jacob G,  Deg\"unther C,
Dawkins S T,  Poschinger U G,  Nigmatullin R,  Retzker A, Plenio M B,
Schmidt-Kaler F, Singer K  2012 {\it Nat. Commun.} {\bf  4} 2290 

\bibitem{EH13} Ejtemaee S and Haljan P C 2013 {\it Phys. Rev. A} {\bf 87} 051401(R)

\bibitem{Esslinger07} Donner T, Ritter S, Bourdel T, \"Ottl A, K\"ohl M, Esslinger T 2007 {\it Science} {\bf 315} 15156
\bibitem{Anderson08} Weiler C N,  Neely T W,  Scherer D R, Bradley A S, Davis M J and  Anderson B P 2008
 {\it Nature} {\bf 455} 948 
 
\bibitem{Anderson07} Scherer D R, Weiler C N, Neely T W, andAnderson B P 2007 {\it Phys. Rev. Lett. } {\bf 98} 110402

\bibitem{Andersonpra08}  Carretero-Gonz\'alez R, Anderson B P, Kevrekidis P G, Frantzeskakis D J and Weiler C N, 2008 {\it Phys. Rev.} A {\bf 77} 033625

\bibitem{DZ10} Damski B and Zurek W H 2010
{\it Phys. Rev. Lett.} {\bf 104} 160404 


\bibitem{WDGR11} Witkowska E.  Deuar P, Gajda M and Rza\.zewski K 2011
{\it Phys. Rev. Lett.} {\bf106} 135301

\bibitem{Das12}Das A, Sabbatini J, and Zurek W H 2012 {\it Sci. Rep.} {\bf 2} 352

\bibitem{newexp}Lamporesi G, Donadello S,  Serafini S,  Dalfovo F,  Ferrari G 2013 arXiv:1306.4523

\bibitem{SZD11} Sabbatini J,  Zurek W H and  Davis M J 2011 
{\it Phys. Rev. Lett.} {\bf 107} 230402 

\bibitem{SKU13} Saito H, Kawaguchi Y and Ueda M 2013 arXiv:1301.4770

\bibitem{DeMarco11} Chen D, White M, Borries C and DeMarco B 2011 
{\it Phys. Rev. Lett.} {\bf 106} 235304  


\bibitem{CK10} Collura M and Karevski D 2010 
Critical Quench Dynamics in Confined Systems.
 {\it Phys. Rev. Lett.} {\bf 104} 200601


\bibitem{CV09} Campostrini M and Vicari E 
2009
 {\it Phys. Rev. Lett.} {\bf 102} 240601







\bibitem{QS82}  Feynman,  R. P. (1982). Simulating physics with computers. {\it Int. J. Theo. Phys.} {\bf 21}, 467.

\bibitem{CZ12}  Cirac J I and Zoller P 2012
{\it Nature Phys.} {\bf 8} 264 


\bibitem{AQC}  Farhi, E,  Goldstone J,  Gutmann S,  Sipser M 2000
  arXiv:quant-ph/0001106

\bibitem{Murg04} Murg V, Cirac J I 2004
 {\it Phys. Rev. } A {\bf 69} 042320

\bibitem{BP08} Barankov R, Polkovnikov A  2008 
{\it Phys. Rev. Lett.} {\bf 101} 076801 

\bibitem{SSM08} Sen D, Sengupta K, Mondal S 2008 
{\it  Phys. Rev. Lett.}  {\bf 101} 016806. 

\bibitem{Patane08} Patan\`e D,  Silva A, Amico L, Rosario  F, Santoro G E 2008 
 {\it Phys. Rev. Lett.}  {\bf 101} 175701 

\bibitem{Bermudez09} Bermudez A, Patan\`e D, Amico L, Martin-Delgado M A 2009
 {\it Phys. Rev. Lett.}  {\bf 102} 135702 

\bibitem{Bermudez10} Bermudez A, Amico L, Martin-Delgado M A 2010 
{\it New J. Phys.} {\bf 12} 055014 

\bibitem{OQC} Doria P,  Calarco T,  Montangero S 2011
{\it Phys. Rev. Lett.}  {\bf 106} 190501 

\bibitem{OQC2} Caneva T,  Calarco T, Fazio R,   Santoro  G E, Montangero S 2011 
{\it Phys. Rev. A} {\bf  84} 012312

\bibitem{OQC3} Nehrkorn J,  Montangero S, Ekert  A, Smerzi A, Fazio R, Calarco T 2011 
arXiv:1105.1707

\bibitem{RC11} Rahmani A and Chamon C 2011 
{\it  Phys. Rev. Lett.}  {\bf 107} 016402 

\bibitem{DRZ12} del Campo A, Rams M M, Zurek W H 2012 
{\it Phys. Rev. Lett} {\bf 109} 115703

\bibitem{GE12} Gammelmark S and Eckardt A  2013 {\it New. J. Phys.} {\bf 15} 033028
\end{thebibliography}
\end{document}